\documentclass[prb,aps,nofootinbib, twocolumn, amssymb,superscriptaddress,longbibliography,10pt]{revtex4-2}
\usepackage{amsmath}
\usepackage{amssymb}
\usepackage{amsthm}
\usepackage{amsfonts}%
\usepackage{enumerate}
\usepackage{latexsym}
\usepackage{color}
\usepackage{setspace} 
\usepackage{blindtext}
\usepackage{dsfont}
\usepackage{mathrsfs}
\usepackage{mathptmx} 
\usepackage[normalem]{ulem}

\usepackage{tabularx}
\newcolumntype{Y}{>{\centering\arraybackslash}X}

\usepackage{stackengine}

\usepackage{bm}
\usepackage{graphicx}
\usepackage{subfigure}

\usepackage{hyperref}
\hypersetup{
pdfnewwindow=true, colorlinks=true,
linkcolor=blue, anchorcolor=blue,
citecolor=blue, filecolor=blue,
menucolor=blue, urlcolor=blue} 

\usepackage{pifont}

\newcommand{\beginsupplement}{
        \setcounter{table}{0}
        \renewcommand{\thetable}{S\arabic{table}}
        \setcounter{figure}{0}
        \renewcommand{\thefigure}{S\arabic{figure}}
        \setcounter{equation}{0}
        \renewcommand{\theequation}{S\arabic{equation}}
        \setcounter{section}{0}
        \renewcommand{\thesection}{\Roman{section}}
        \renewcommand{\thesubsection}{\arabic{subsection}}
}

\newcommand{\vk}{\mathbf{k}}

\begin{document}

\title{Electron correlations in the kagome flat band metal \texorpdfstring{$\rm \bf CsCr_3Sb_5$}{CsCr3Sb5}}

\author{Fang Xie}
\affiliation{Department of Physics \& Astronomy,  Rice Center for Quantum Materials, Rice University, Houston, Texas 77005, USA}
\affiliation{Rice Academy of Fellows, Rice University, Houston, Texas 77005, USA}
\author{Yuan Fang}
\affiliation{Department of Physics \& Astronomy,  Rice Center for Quantum Materials, Rice University, Houston, Texas 77005, USA}
\author{Ying Li}
\affiliation{MOE Key Laboratory for Nonequilibrium Synthesis and Modulation of Condensed Matter, School of Physics, Xi'an Jiaotong University, Xi'an 710049, China}
\author{Yuefei Huang}
\affiliation{Department of Materials Science and NanoEngineering, Rice University, Houston, Texas 77005, USA}
\author{Lei Chen}
\affiliation{Department of Physics \& Astronomy,  Rice Center for Quantum Materials, Rice University, Houston, Texas 77005, USA}
\author{Chandan Setty}
\affiliation{Department of Physics \& Astronomy,  Rice Center for Quantum Materials, Rice University, Houston, Texas 77005, USA}
\affiliation{Department of Physics and Astronomy, Iowa State University, Ames, Iowa 50011, USA}
\affiliation{Ames National Laboratory, U.S. Department of Energy, Ames, Iowa 50011, USA}
\author{Shouvik Sur}
\affiliation{Department of Physics \& Astronomy,  Rice Center for Quantum Materials, Rice University, Houston, Texas 77005, USA}
\author{Boris Yakobson}
\affiliation{Department of Materials Science and NanoEngineering, Rice University, Houston, Texas 77005, USA}
\author{Roser Valent\'\i}
\affiliation{Institut f\"ur Theoretische Physik, Goethe-Universit\"at Frankfurt, Max-von-Laue-Strasse 1, 60438 Frankfurt am Main, Germany}
\author{Qimiao Si}
\affiliation{Department of Physics \& Astronomy,  Rice Center for Quantum Materials, Rice University, Houston, Texas 77005, USA}

\date{\today}
\begin{abstract}

Kagome metals offer a unique platform for investigating robust electron-correlation effects because of their lattice geometry, flat bands and multi-orbital nature. 
In the cases with active flat bands,  recent theoretical studies have pointed to a rich  phase diagram that contains not only electronic orders but also quantum criticality. 
Very recently, $\rm CsCr_3Sb_5$ has emerged as a strong candidate for exploring such new physics.
Here, using effective tight-binding models obtained from ab initio calculations, we study the effects of electronic correlations and symmetries on the electronic structure of $\rm CsCr_3 Sb_5$. The effective tight-binding model and Fermi surface comprise multiple Cr-$d$ orbitals and Sb-$p$ orbitals. The introduction of Hubbard-Kanamori interactions leads to orbital-selective band renormalization dominated by the $d_{xz}$ band, concurrently producing emergent flat bands very close to the Fermi level. Our analysis sets the stage for further investigations into the electronic properties of $\rm CsCr_3Sb_5$, including electronic orders, quantum criticality and unconventional superconductivity, which promise to shed much new light into the electronic materials with frustrated lattices and bring about new connections with the correlation physics of a variety of strongly correlated systems.  

\end{abstract}

\maketitle

{\it Introduction.}
Materials with flat bands from frustrated lattices provide a distinctive venue to explore the intersection of lattice geometry, strong correlation effects and multi-orbital physics \cite{Mielke_1992, Lin2023Complex, Calugaru2022, Hasan2020, Comin2020-2, Guguchia2021-TRSB, Zhou2022, Dai2022, Hasan2022-PRL, Setty2022}. 
Experiments in such materials have found evidence for non-Fermi liquid behavior \cite{Yi2024-CVS,Ye2024,Ekahana2024}.
Theoretically, the coupling between the flat and wide bands has been shown to yield a formation of local moment from compact molecular orbitals \cite{Si2023-TopologyQCP, Chen_emergent_2024, hu2023coupled} and the development of electronic orders, with the quantum critical regime displaying strange metal behavior \cite{Si2023-TopologyQCP, hu2023coupled}. 
The recent discovery of $\rm CsCr_3Sb_5$~\cite{Liu2024Superconductivity}, a Cr-based kagome material, has opened an exciting avenue~\cite{sangiovanni2024superconductor} for exploring this rich physics while adding a {\it distinct} new member to the kagome material family~\cite{JXYin2022-Review}.  With its unique phenomenology and multi-orbital nature, this system offers a rare opportunity to examine the robust effects of electron correlation in a flat-band platform.

$\rm CsCr_3Sb_5$ is isostructural to the more familiar vanadium-based kagome material family $A\rm V_3Sb_5$ ($ A = \rm Cs, K, Rb$)~\cite{Ortiz2019, Ortiz2020, Ortiz2021, Wilson2023-Review}. It is a bad metal at ambient pressure supporting the existence of strong correlation effects. Anomalies in transport and thermodynamic properties suggest multiple ordering transitions as a function of temperature and pressure~\cite{Liu2024Superconductivity}. 
While the nature of the order is presently unclear, it has a single ordering wave-vector modulation, involving a hexagonal to monoclinic distortion~\cite{Liu2024Superconductivity}.
This is different from the behavior observed in the vanadium family~\cite{Hasan2020, Zeljkovic2021, Wen2021, XHChen2021, Tsirlin2021, Miao2021, Shi2021}.

Equally important, high temperature susceptibility data in $\rm CsCr_3Sb_5$ shows a Curie-Weiss behavior indicating the formation of local moments.
High pressure suppresses the density-wave ordering, leading to dome-like unconventional superconductivity separated by a putative quantum critical point; in contrast, the $A\rm V_3Sb_5$ family exhibits ambient pressure superconductivity, which persists under pressure~\cite{Wilson2023-Review}. In addition, resistivity data indicate strong deviations from Fermi-liquid behavior further supporting the role of strong correlations and quantum criticality~\cite{Liu2024Superconductivity}. 

Here, we study the low-energy correlated electronic properties of $\rm CsCr_3Sb_5$, including its flat band behavior near the Fermi energy and Fermi surface characteristics that shape the physical properties of this material; importantly, we rest our understanding of the low-energy electronic states on a comprehensive symmetry analysis.
We clarify the role of strong correlation effects on the electronic structure by investigating the suppression of the orbital-dependent quasiparticle weights using the slave-spin approach. 
We begin by identifying the site-symmetry groups and representations of Wyckoff positions of the space group $P6/mmm$, which correspond to the lattice sites of $\rm Cr$ and $\rm Sb$ atoms.
We then analyze the band structure obtained from {\it ab initio} calculations, which identify the orbitals from the Cr and Sb atoms that are relevant to the low-energy physics and lead to a tight-binding model.
We find that the Hubbard-Kanamori-type interactions~\cite{Kanamori1963, Georges2013Strong} induce strong
orbital-selective renormalization effects in the Cr-$d$ orbitals, especially the $d_{xz}$ orbital.
Additionally, our analysis reveals that strong correlations lead to a  noticeably larger spectral weight near the Fermi level, in contrast to the prediction of {\it ab initio} study. 
In other words, the correlations generate an emergent flat band very close to the Fermi energy.
Our investigation sets the stage for further research into the nature of the
unusual electronic order, quantum criticality and superconductivity of this promising new material.

\begin{figure}[t]
    \centering
    \includegraphics[width=\linewidth]{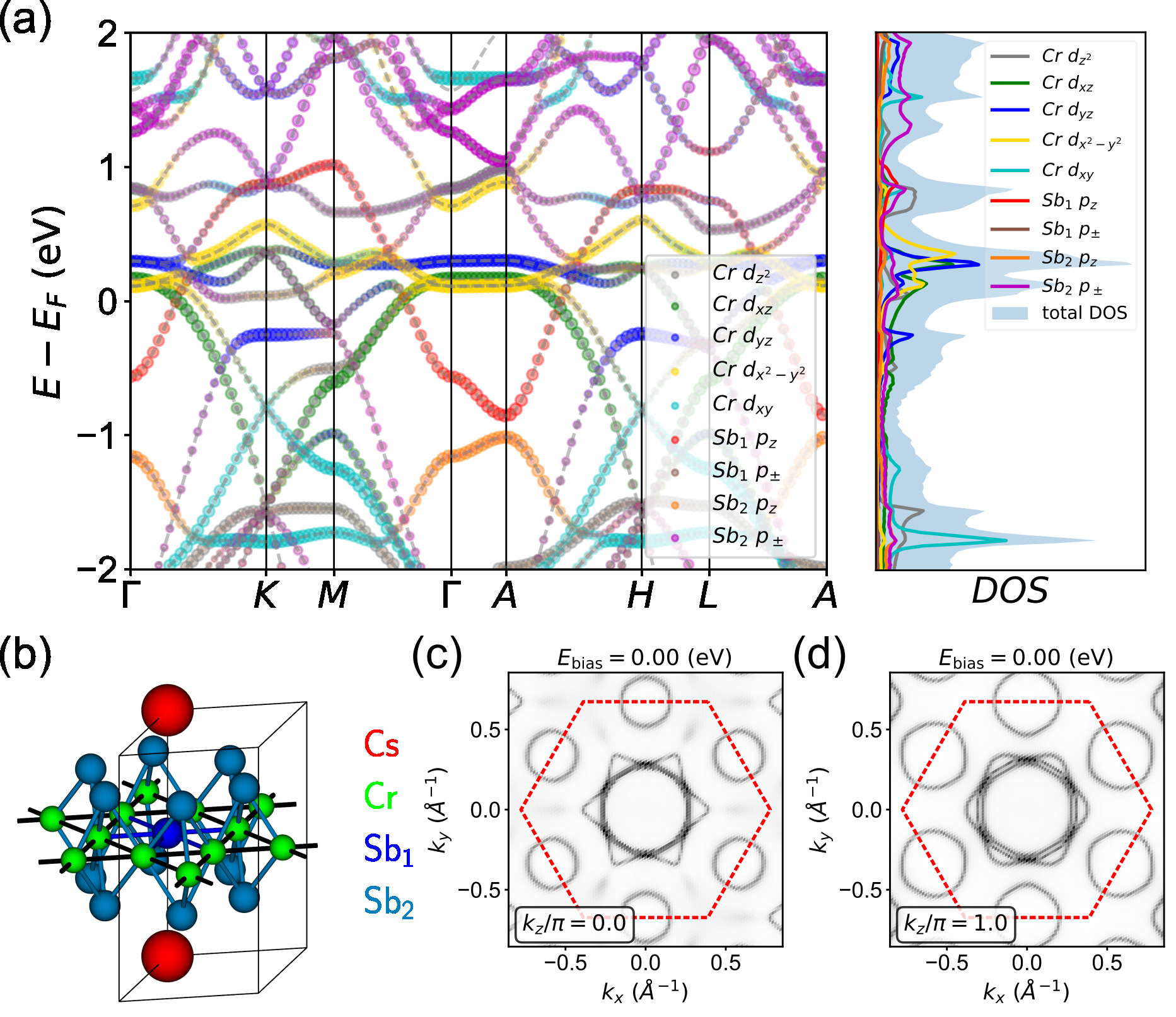}
    \caption{
    (a) Band structure and density of states of $\rm CsCr_3Sb_5$ compound obtained from density functional theory using VASP and GGA. The color codes represent the orbital content of the corresponding Bloch states.
    (b) The lattice structure of $\rm CsCr_3Sb_5$. Black box stands for the unit cell.
    (c-d) Momentum-resolved spectral function at $k_z = 0$ and $k_z = \pi$ planes at the Fermi energy obtained from the {\it ab initio} band structure. The red dashed line stands for the first Brillouin zone. Electron pockets near the $M$ points, electron \& hole pockets around $\Gamma$ point can be seen in this spectral function.
    Spin-orbital-coupling (SOC) is not considered.
    }
    \label{fig:dft}
\end{figure}

\begin{figure}
    \centering
    \includegraphics[width=\linewidth]{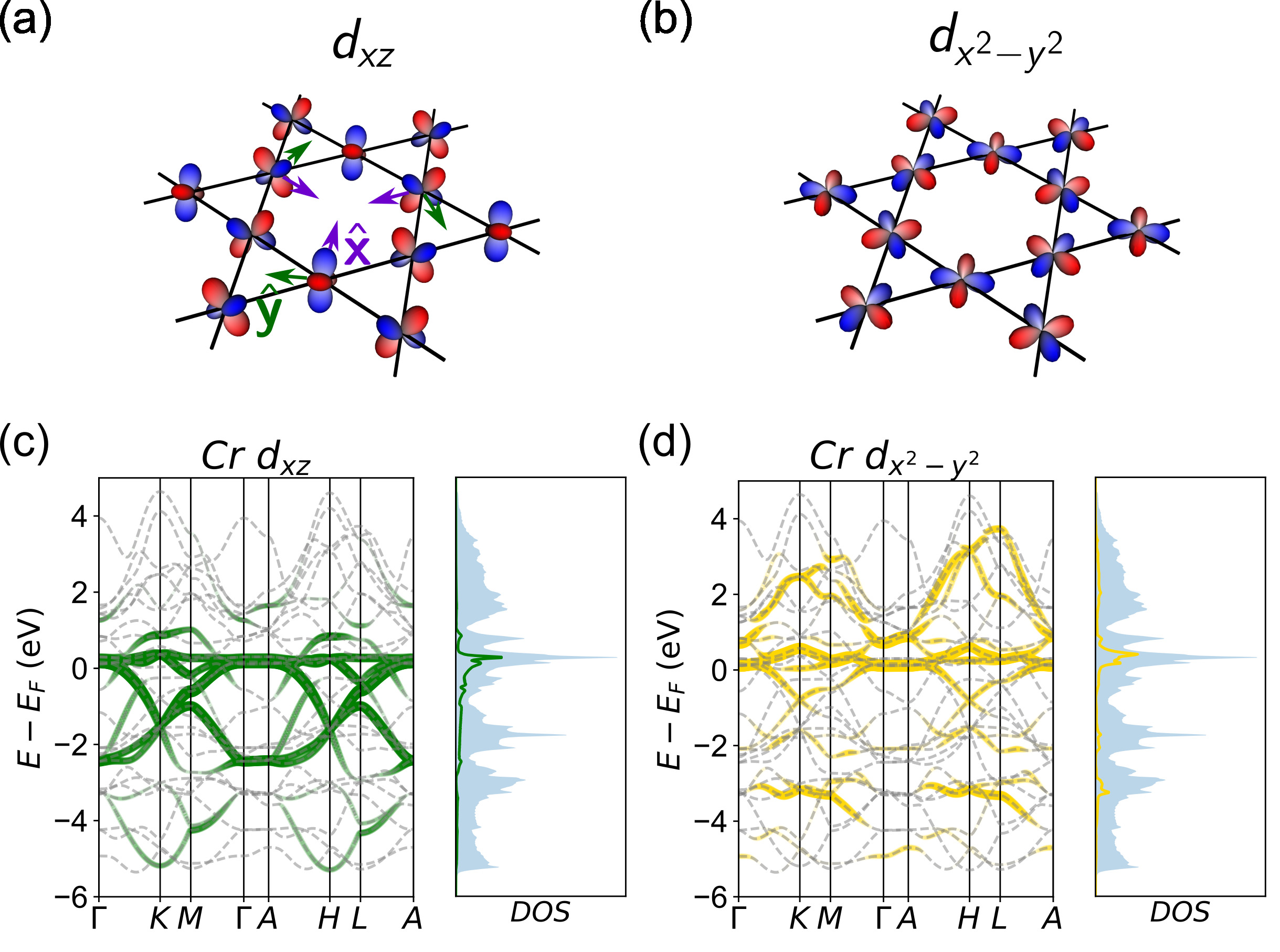}
    \caption{(a) The shape of the $d_{xz}$ orbitals on the chromium atoms. The basis vectors $\hat{\mathbf{x}}, \hat{\mathbf{y}}$ of local coordinate frame on each kagome site are represented by the purple and green arrows, respectively. Note that the orientation of the principle axes on different kagome sites are chosen differently. (b) The shape of the $d_{x^2 - y^2}$ orbitals on the chromium atoms. 
    (c) Band structure and the density of states (DOS) projected onto the $d_{xz}$ orbitals.
    (d) Band structure and the DOS projected onto the $d_{x^2 - y^2}$ orbitals.
    In the DOS plots, the light blue shadows stand for the total DOS.
    }
    \label{fig:orbs}
\end{figure}

\begin{table}[t]
    \centering
    \begin{tabular}{cccc|c|c}
        \hline
         $E$ &$C_{2x}$ &$C_{2y}$ &$C_{2z}$ &irrep &orbital  \\
         \hline
         $+1$ &$+1$ &$+1$ &$+1$ &$A_g$ & $d_{z^2}$, $d_{x^2-y^2}$\\
         $+1$ &$-1$ &$-1$ &$+1$ &$B_{1g}$ &$d_{xy}$\\
         $+1$ &$-1$ &$+1$ &$-1$ &$B_{2g}$ &$d_{xz}$\\
         $+1$ &$+1$ &$-1$ &$-1$ &$B_{3g}$ &$d_{yz}$\\
         \hline
    \end{tabular}
    \caption{Irreducible representations of $d$-orbitals in the point group $mmm$ ($D_{2h}$) of the $3f$ Wyckoff position of space group 191. Since all $d$-orbitals have inversion eigenvalue $+1$, the three two-fold rotations distinguish their irreps.}
    \label{tab:pg-irrep}
\end{table}

{\it Symmetry and effective model.}
The $\rm CsCr_3Sb_5$ compound has nine atoms per unit cell. As shown in Fig.~\ref{fig:dft}(b), there are three Cr atoms forming a kagome lattice. 
Among the five Sb atoms, one of them is at the center of the hexagon plaquette of the chromium kagome lattice within the Cr plane denoted as $\rm Sb_1$, while the other four are at out-of-plane positions denoted as $\rm Sb_2$, forming two sets of hexagonal lattices. 
Besides, the Cs atom is positioned atop of $\rm Sb_1$ in a separate plane.

The active orbitals near the Fermi level are the Cr-$3d$ orbitals, and the Sb-$5p$ orbitals.
Indeed, these orbitals are all responsible for the low energy physics near the Fermi level, as suggested by {\it ab initio} studies \cite{Liu2024Superconductivity, Xu2023Frustrated}. 
In Fig.~\ref{fig:dft}(a) we provide the band dispersion, whose color codes stand for the weights of different orbital content, and the corresponding density of states (DOS) obtained from density functional theory using VASP \cite{Kresse1996Efficient, Kresse1999From} and GGA \cite{Perdew1996GGA}. The total density of states approaches its peak at around $\sim 0.3\rm\, eV$ above the Fermi level, which are mostly formed by the Cr-$d_{xz}$ and $d_{yz}$ orbitals. The $d_{z^2}$ and $d_{x^2-y^2}$ orbitals are also responsible for states close to $E_F$. 
Except for these Cr orbitals, we also found that the $\rm Sb_1$-$p_z$ orbital provides an electron pocket near the $\Gamma$ point, which is highlighted by the red markers in Fig.~\ref{fig:dft}(a). 
The $\rm Sb_2$-$p_\pm$ and $p_z$ orbitals also strongly hybridize with the Cr-$d$ orbitals, which lead to non-negligible contributions to the band structure near $E_F$. 

We show the momentum resolved spectral function at the Fermi level with different $k_z$ values obtained from the {\it ab initio} band structure in Figs.~\ref{fig:dft}(c) and \ref{fig:dft}(d).
Three electron pockets near the $M$ points, which are mostly formed by Cr-$d_{xz}$ and $d_{yz}$ orbitals, are easily visible. Furthermore, we observe near the $\Gamma$ point one electron pocket formed by the Sb-$p_z$ orbital, and hole pockets formed by  Cr-$d_{xz}$, $d_{z^2}$ and $d_{x^2-y^2}$ orbitals. The similarity in the Fermi pockets at $k_z = 0$ and $k_z = \pi$ planes also shows the two-dimensional nature of this material.

This {\it ab initio} analysis delineates the degrees of freedom that are important for the electronic structure of $\rm CsCr_3Sb_5$.
Since there are many orbitals contributing to the low energy physics, it is a challenging task to build a reliable model to capture the relevant physics while not being unnecessarily complicated. 

To make progress, we first analyse the crystalline symmetry.
The space group of $\rm CsCr_3Sb_5$ is $P6/mmm$ (no.~191). The Wyckoff positions, site-symmetry groups and irreducible representations (irrep) of these active atomic orbitals are discussed as follows \cite{Bradlyn2017Topological, Elcoro2017double}:
\begin{itemize}

\item 
The Cr-$d$ orbitals are at the $3f$ Wyckoff position (kagome sites). Each of the orbitals forms a one-dimensional irrep of the site-symmetry group $mmm$ ($D_{2h}$). 
Since all $d$ orbitals have inversion eigenvalue $+1$, their representations are uniquely determined by the quotient group $222$ ($D_2$).
In Table \ref{tab:pg-irrep}, we provide the character table and the representations of the $d$-orbitals of this point group. 
The local coordinate basis for each kagome site is selected in accordance with the principle axes of the site symmetry group, as labeled in Fig.~\ref{fig:orbs}(a).

\item 
The $\rm Sb_1$-$p_z$ orbital is at the $1b$ Wyckoff position (triangular sites) which forms the irrep $A_{2u}$ of the site-symmetry group $6/mmm$ ($D_{6h}$). The $\rm Sb_1$-$p_{\pm}$ orbitals form the irrep $E_{1u}$ of the site-symmetry group $6/mmm$.

\item 
The $\rm Sb_2$-$p_z$ and $p_\pm$ orbitals are at the $4h$ Wyckoff position which form the one-dimensional irrep $A_1$ and two-dimensional irrep $E$ of the site-symmetry group $3m$ ($C_{3v}$), respectively.

\item 
One empty Cs band is around $2\rm\,eV$ above the Fermi level, which also hybridizes with multiple Sb-$p$ orbitals. The Cs-$s$ orbital is at the $1a$ Wyckoff position which forms the irrep $A_{1g}$ of the site-symmetry group $6/mmm$ ($D_{6h}$). 
    
\end{itemize}

By projecting the {\it ab initio} Bloch states onto these atomic orbitals, one can construct an effective tight-binding model with $31= 1 \times 1 (\textrm{Cs-}s) + 3\times 5 (\textrm{Cr-}d) + 5 \times 3(\textrm{Sb-}p)$ orbitals and $34 = 1 \times 1 (\textrm{Cs-}6s^1) + 3 \times 6 (\textrm{Cr-}3d^54s^1) + 5\times 3(\textrm{Sb-}5p^3)$ valence electrons per unit cell using \textsc{Wannier90} \cite{Pizzi2020wannier}. 
Note that all of the five $d$ orbitals on each of the Cr atom are taken into account in this tight-binding model. 
As such, this tight-binding model can faithfully describe the energy bands in the interval $ -6\,{\rm eV} < E - E_F < 5\rm\,eV$. 

In Figs.~\ref{fig:orbs}(c-d), we provide the ``fat bands'' projected onto the $d_{xz}$ and $d_{x^2 - y^2}$ orbitals on the 31-band tight-binding model, with their corresponding DOS plots.
The ``fat bands'' plots of the other three types of $d$ orbitals are shown in Fig.~S1 of the supplemental material (SM) \cite{supplemental_material}.
Due to the strong hybridization between the Cr-$d$ orbitals and the Sb-$p$ orbitals, states that are a few electron-volts away from the Fermi level still possess non-negligible DOS from these Cr-$d$ orbitals.
Among the five $d$ orbitals, we found that the DOS of the $d_{xz}$ orbital has the smallest spread, while those of the $d_{yz}$ and $d_{x^2 - y^2}$ orbitals have larger spreads than for the other orbitals.
These results indicate that the $d_{xz}$ orbital has a weaker hybridization strength with the other orbitals, suggesting that it will have the strongest renormalization when the interaction effects are considered.
In addition, it is also noticeable that all of these five Cr-$d$ orbitals are relevant to the electronic states near the Fermi level.
Hence, any truncated effective models with a subset of these orbitals on Cr atoms are unlikely to be faithful.

{\it Interaction effects.}
As addressed in the previous section, an effective model with truncated Cr-$d$ orbitals is not suitable for accurately describing correlation effects in low-energy physics. 
Therefore, a proper treatment of interaction terms should include all these Cr-$d$ orbitals, with the Cs-$s$ and Sb-$p$ orbitals as the ``conduction bath'' states.
We choose a Hubbard-Kanamori-type Hamiltonian to describe  the correlation effects \cite{Kanamori1963,Georges2013Strong}, which contains the on-site repulsive interaction and Ising-type ferromagnetic Hund's coupling terms for all five Cr-$d$ orbitals:
\begin{align}
    H_I =& U\sum_{j\alpha}\tilde{n}_{j\alpha\uparrow}\tilde{n}_{j\alpha\downarrow} + U' \sum_{j, \alpha\neq\beta}\tilde{n}_{j \alpha\uparrow}\tilde{n}_{j\beta\downarrow} \nonumber\\
&    + (U' - J_H)\sum_{j, \alpha < \beta, \sigma}\tilde{n}_{j\alpha\sigma}\tilde{n}_{j \beta\sigma}\,.\label{eqn:hubbard-kanamori-hamiltonian}
\end{align}
Here $j$ stands for the kagome lattice sites, $\alpha, \beta = 1,2,\cdots, 5$ mark the Cr-$d$ orbital indices, and $\sigma = \uparrow,\downarrow$ describes the electron spin. 
The operator $\tilde{n}_{j,\alpha,\sigma} = c^\dagger_{j,\alpha,\sigma}c_{j,\alpha,\sigma} - \frac12$ represents the (relative) electron density on orbital $\alpha$ with spin $\sigma$ at site $j$. 
The bare on-site Coulomb interaction on the Cr-$d$ orbitals is estimated to be around $U_{\rm bare} \approx 12 \sim 15\,\rm eV$. However, screening effects and the type of chemical environment of Cr reduce the effective interaction strength to values around $3\sim 6\,\rm eV$ \cite{Soriano2021, Vaugier2012Hubbard}.

The interacting Hamiltonian in the correlated metallic phases can be studied by the $U(1)$ slave-spin approach \cite{Yu2012U1slave, yu_mott_2011, supplemental_material}. 
In this framework, the quasiparticle weights and the on-site potential renormalization for each orbital can be solved self-consistently \cite{Lin-slave-spin-z-zprime2024}.
The meaning of these renormalization factors is two-fold: 
(i) the quasiparticle weights which are smaller than unity indicate the suppression of the kinetic energies and the effective hybridization strength in the coherent Fermi liquid excitations;
(ii) the on-site potential renormalizations may lead to charge transfer among different orbitals, while they guarantee the total filling factor to be unchanged when the interaction strength is varied. With both renormalization factors considered, the dispersion of the coherent Fermi liquid excitations can be solved. In addition, the shape and the orbital contents of the Fermi surface may vary when the interactions are taken into consideration, due to orbital-selective nature of the correlation effects.

\begin{figure}[t]
    \centering
    \includegraphics[width=\linewidth]{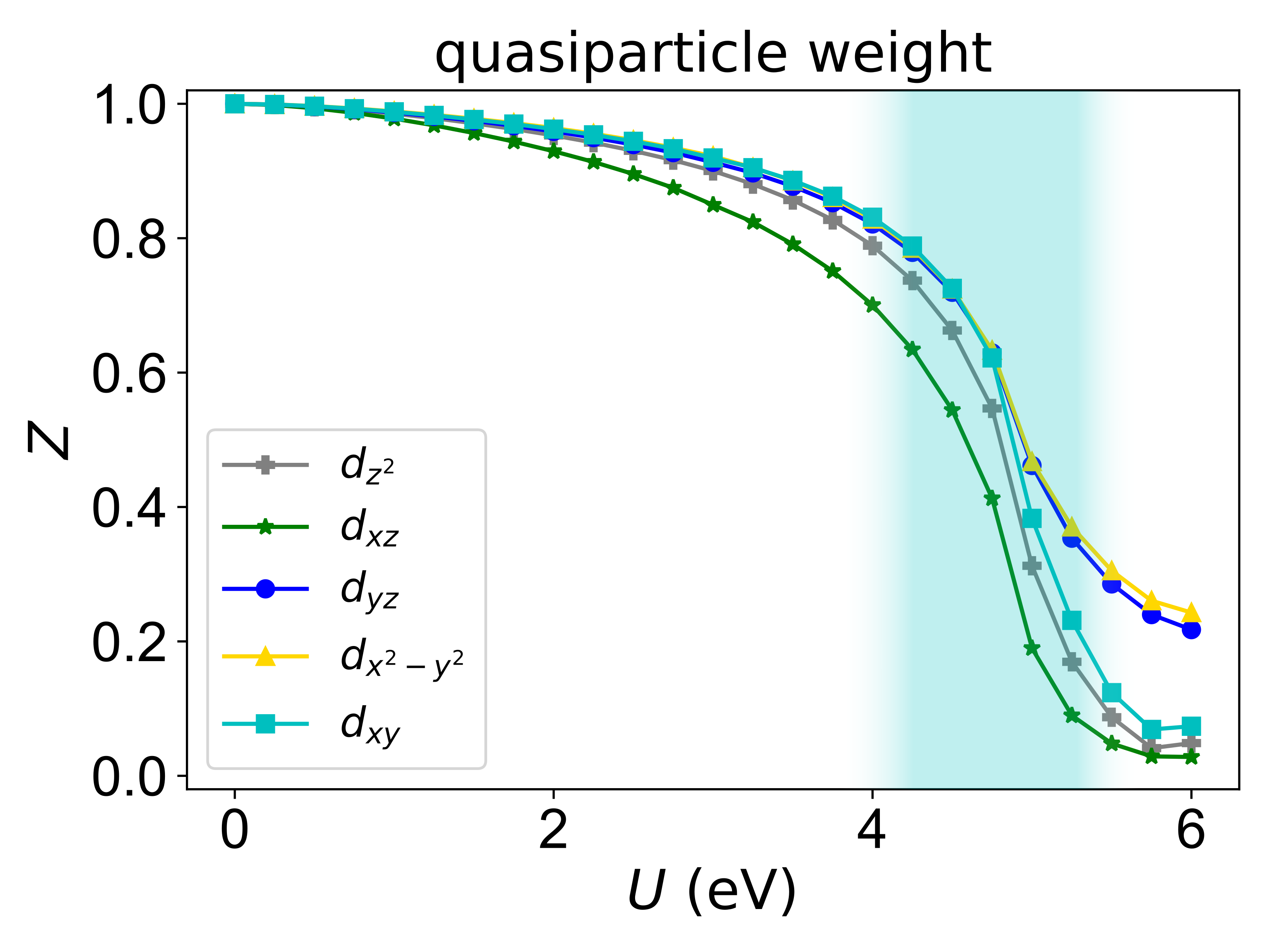}
    \caption{The quasiparticle weights and electron occupation numbers (per spin) of the five Cr-$d$ orbitals, computed as functions of the interaction strength $0 \leq U \leq 6\,{\rm eV}$. The $d_{xz}$ orbital shows the strongest renormalization effect with large interaction strength $U \gtrsim 4\rm\,eV$. In the region highlighted by the cyan shadow, the $d$ orbitals' quasiparticle weights drop significantly.
    The Hund's coupling strength is chosen to be $J_H = 0.2 U$ in this calculation.
    }
    \label{fig:phase-diag}
\end{figure}

\begin{figure}[t]
    \centering
    \includegraphics[width=\linewidth]{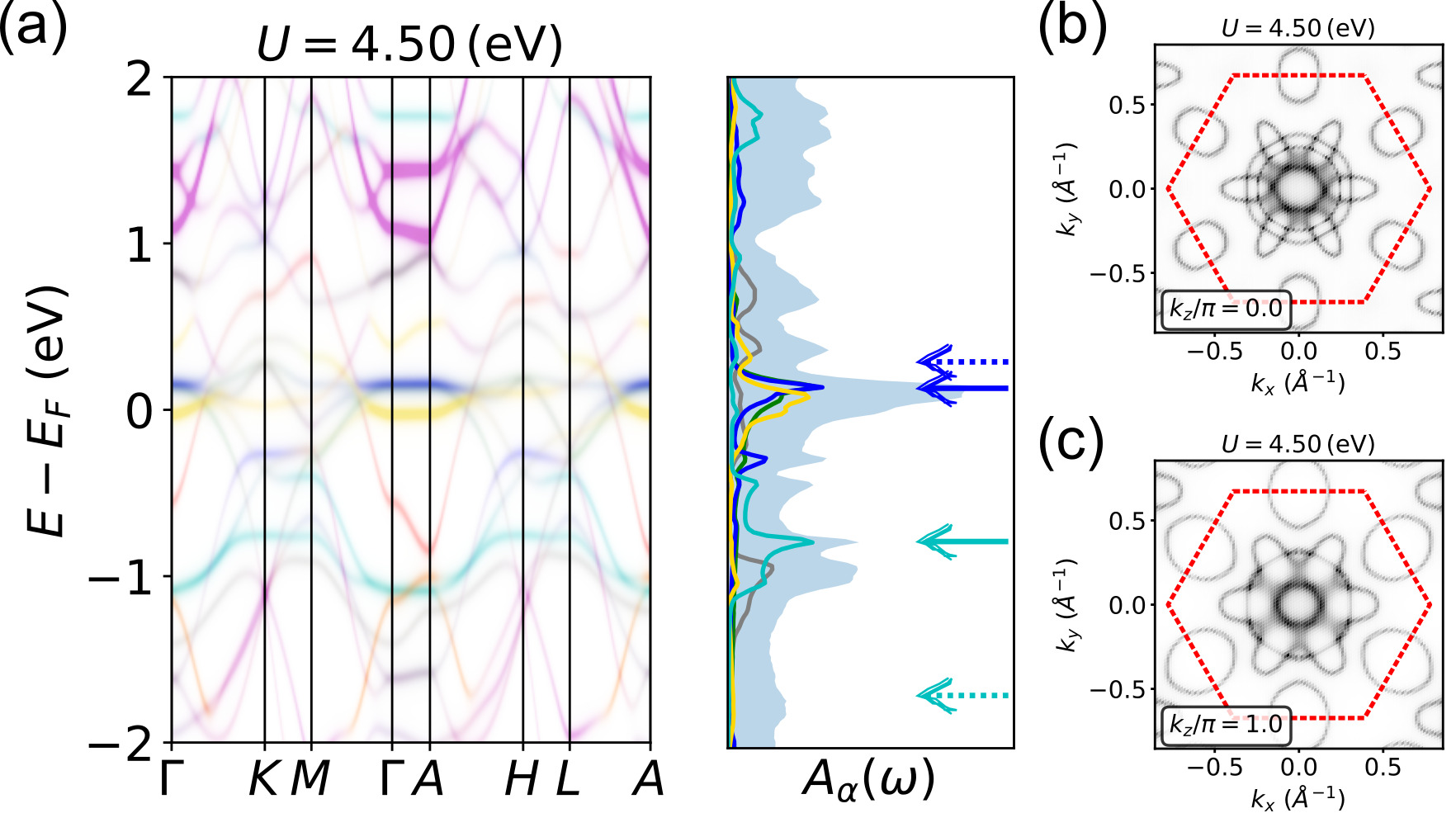}
    \caption{(a) Momentum resolved spectral functions of the coherent excitations, $A_\alpha^{\rm coh}(\vk, \omega)$, and the density of states obtained from the slave-spin approach with the interaction strength $U = 4.5\rm\,eV$. 
    The color codes are the same as labeled in Fig.~\ref{fig:dft}. 
    The dashed arrows stand for the peaks of DOS with $U = 0$, while the solid arrows mark the peaks of DOS with $U = 4.5\,\rm eV$.
    The color codes of the orbital contents are the same as in Fig.~\ref{fig:dft}(a).
    (b-c) Fermi surface plots in the $k_z = 0$ and $k_z = \pi$ planes.
    }
    \label{fig:interaction-bands}
\end{figure}

Here we solve the interacting Hamiltonian with 
interaction strength $0 \leq U \leq 6\,{\rm eV}$ and Hund's coupling $J_H/U = 0.2$ at a fixed filling factor. 
Our results indicate that: (i) the orbital-selective correlation effects exist, and the $d_{xz}$ orbital experiences the strongest renormalization; (ii) the narrow bands are pinned near the Fermi energy, leading to a considerable reduction in the energetic separation between the Fermi level and the peak of the DOS from the {\it ab initio} value of $0.3\rm\, eV$. Below, we further elaborate on our discussion regarding these points.

{\it Orbital-selective Mott correlation.}
The quasiparticle weights of all the five $d$ orbitals as functions of the interaction strength $U$ are provided in Fig.~(\ref{fig:phase-diag}). 
Significantly, the quasiparticle weights are strongly renormalized when the interaction strength $U$ is increased above $4\rm\,eV$.
Moreover, the renormalization effects on distinct $d$ orbitals are different from each other.
In brief, the $d$ orbitals which have weaker hybridization or smaller hoppings to other orbitals are more likely to be renormalized. 
Indeed, as shown in Fig.~\ref{fig:phase-diag}, the quasiparticle weight of the $d_{xz}$ orbital drops the most rapidly with increasing interaction $U$. 
In contrast, the $Z_\alpha$ values of the $d_{x^2 - y^2}$ and $d_{yz}$ orbitals are also noticeably larger than for other $d$ orbitals, especially when $U \gtrsim 5\rm\,eV$. 
Additionally, we observe that if $U$ is increased to $\sim 6\,\rm eV$, which is still within a reasonable range~\cite{Soriano2021,Vaugier2012Hubbard}, the three orbitals ($d_{xz}, d_{z^2}$ and $d_{xy}$) are strongly renormalized with quasiparticle weights $Z \lesssim 0.1$.
This indicates that the system is approaching an orbital-selective Mott transition, during which the quasiparticle weights of one or more orbitals decay to zero.

{\it Emergent flat band near Fermi level.}
Correlation effects on the quasiparticle weights and on-site potentials lead to the suppression of the kinetic energy and hybridization effects in the coherent Fermi liquid spectrum. 
In Fig.~\ref{fig:interaction-bands}(a), the orbital-resolved coherent fermion excitation spectrum $A^{\rm coh}_\alpha(\vk, \omega)$ and its corresponding DOS with interaction strength $U = 4.5\rm\, eV$ are provided. 
The band structure is drastically changed compared to its noninteracting counterpart [Fig.~\ref{fig:dft}(a)]. 
The dashed arrows in Fig.~\ref{fig:interaction-bands}(a) label the position of two DOS peaks in the {\it ab initio} band structure, while the solid arrows are the corresponding peaks in the correlated DOS.
The peak above $E_F$ labeled by the blue arrow, which is mostly formed by $d_{xz}, d_{yz}$ and $d_{x^2 - y^2}$ orbitals, now appears considerably closer to the Fermi energy.
At the same time, the occupied ``flat band'' of the $d_{xy}$ orbital, which is highlighted by the cyan arrows, is strongly pushed upward.
Although the emergent ``flat bands'' of the coherent excitations appear near $E_F$, the reduction of the quasiparticle weights also leads to weaker intensities in the momentum resolved spectral functions.
We note that the correlated flat bands still capture the destructive interference of the kagome lattice, since they can be adiabatically connected to the destructive-interference-induced flat bands in the DFT bandstructure 
when the interaction $U$ is gradually reduced; this point is further discussed in the SM \cite{supplemental_material}.

The correlated Fermi surfaces are also shown in Figs.~\ref{fig:interaction-bands}(b) and \ref{fig:interaction-bands}(c). 
Due to the orbital-selective renormalization effects, the size and shape of Fermi pockets with different orbital contents will be changed. For example, the electron pockets around the three $M$ points become smaller than the ones in Fig.~\ref{fig:dft}(c). The concentric Fermi pockets around the $\Gamma$ point, which are predominantly formed by $p_z$, $d_{xz}$ and $d_{x^2 - y^2}$ orbitals, also show varying sizes compared to their counterparts in the absence of interactions.
Moreover, another shallow electron pocket emerges around $\Gamma$ point, which is mostly formed by the $d_{x^2 - y^2}$ orbital. 
Additional discussion regarding the correlated Fermi surfaces with different interaction strengths $U$ can be found in SM \cite{supplemental_material}.

{\it Discussion.}
Given that  $\rm CsCr_3 Sb_5$ is a new material and is strongly correlated with active flat bands, and considering the extensive ongoing interest in its more weakly correlated V-based counterpart \cite{Wilson2023-Review}, we can expect this Cr-based kagome material to reveal much insight into the strong correlation physics of metallic systems with geometrically frustrated lattices beyond what one can from their V-based counterparts.
Here we have analyzed its electronic states. 
Based on an effective model that faithfully describes the {\it ab initio} band structure, we are able to systematically study the correlation effects on the Cr-$d$ orbitals.
Our results demonstrate strong renormalization effects when the interaction strength is within a reasonable and realistic range. 
A recent study \cite{Wang2024Heavy} using the $\rm DFT+DMFT$ approach reached compatible conclusions regarding the renormalization effects, 
though it did not address either the Fermi surface under correlation or systematic symmetry analysis.

Due to the inequivalence of the Cr-$d$ orbitals as irreps of the point group $D_{2h}$, the correlation effects show a noticeable orbital-selective signature, with the $d_{xz}$ orbital experiencing the largest renormalization. 
The strong correlation effect suggests that fluctuations in the spin and other collective channels will have considerable spectral weight, and this can be tested by neutron scattering or resonant inelastic X-ray spectroscopy.

Our study also provides predictions about the electronic structure of this material.
We identify flat bands that lie in the immediate vicinity of the Fermi energy, in contrast to the noninteracting flat bands that appear about $0.3\rm\,eV$ away from the Fermi energy.
Furthermore, we determine the shape and size of the Fermi pockets in the presence of interactions.
These effects can be tested by measurements such as angle resolved photoemission spectroscopy and scanning tunneling microscopy.
For example, the emergent flat band and related low-energy electronic properties that are predicted here have been observed by three independent experiments based on angle resolved photoemission spectroscopy \cite{Guo2024Ubiquitous, Li2024Correlated,Peng2024Flat}.

Our construction and analysis of the electronic states based on crystalline symmetries set the stage for further in-depth study of the competing ground states and quantum criticality in this new correlated material. 
The fact that $d_{xz}$ and $d_{yz}$ states dominate the low-energy regime suggests the construction of either corresponding quasi-molecular orbitals \cite{Mazin2012Na2IrO3,foyevtsova2013} or the compact molecular orbitals \cite{Si2023-TopologyQCP, Chen_emergent_2024, hu2023coupled} will have the advantage of incorporating the low-energy $p$ and $d$ states in a symmetry-compatible way and addressing the role of topology.
Such real-space basis can also enable the treatment of competing {\it effective} interactions, responsible for both the development of electronic orders and the amplification of quantum fluctuations.
In other kagome compounds, novel density wave phases that break time-reversal symmetry have been evidenced~\cite{Guguchia2021-TRSB, Zhao2021-TRSB, Wang2021-Kerr, Hasan2022-PRL}; the extent to which such unusual electronic order develops in $\rm CsCr_3 Sb_5$, and how they entwine with the electronic topology~\cite{Setty2021, Setty2022} when considering the spin-orbit coupling \cite{supplemental_material}, are additional fascinating questions for future studies.

\begin{acknowledgments}
We thank Pengcheng Dai, Ming Yi, Yucheng Guo, Zehao Wang and Yiming Wang for useful discussions. 
Work at Rice has primarily been supported  by the U.S. DOE, BES, under Award No. DE-SC0018197, and by the Robert A. Welch Foundation Grant No. C-1411.
C.S. additionally acknowledges support from Iowa State University and Ames National Laboratory startup funds.
The majority of the computational calculations have been performed on the Shared University Grid at Rice funded by NSF under Grant EIA-0216467, a partnership between Rice University, Sun Microsystems, and Sigma Solutions, Inc., the Big-Data Private-Cloud Research Cyberinfrastructure MRI-award funded by NSF under Grant No. CNS-1338099, and the Advanced Cyberinfrastructure Coordination Ecosystem: Services \& Support (ACCESS) by NSF under Grant No. DMR170109. 
Y.L. acknowledges funding by the National Natural Science Foundation of China (Grant No.\ 12004296). 
R.V. acknowledges funding by the Deutsche Forschungsgemeinschaft (DFG, German Research Foundation) through Project No. TRR 288 --- 422213477 (project A05).
The work of Y.H. and B.Y. was supported by the U.S. Office of Naval Research grant N00014-22-1-2753.

\end{acknowledgments}

\bibliography{ref}

\clearpage

\onecolumngrid
\beginsupplement
\section*{Supplemental Material}

\section{Slave-spin approach}\label{appsec:ss}

In this work, we studied the correlated electronic structure of CsCr$_3$Sb$_5$, based on the kinetic Hamiltonian, which has all of the Cr-$d$, Sb-$p$ and Cs-$s$ orbitals in the $31$-band model.
As we have already addressed in the main text, a faithful representation of the electronic structure near the Fermi level require all five $d$ orbitals from the Cr atoms [c.f. Fig.~\ref{fig:orbs_app}].
Therefore, we consider all the 31 relevant orbitals while accounting for the Hubbard-Kanamori interaction terms on all five $d$ orbitals of each Cr atom.

In this section, we briefly review the $U(1)$ slave-spin approach \cite{Yu2012U1slave}.
Generically, the kinetic Hamiltonian has the following tight-binding form:
\begin{equation}
    H_0 = \sum_{\mathbf{k}\alpha\beta\sigma}t_{\alpha\beta}(\vk)c^\dagger_{\vk\alpha\sigma}c_{\vk\beta\sigma} + \sum_{\vk\alpha\sigma}\tilde{\varepsilon}_\alpha c^\dagger_{\vk\alpha\sigma} c_{\vk\alpha\sigma}\,.
\end{equation}
In the context of electronic structure of CsCr$_3$Sb$_5$, the orbital indices $\alpha,\beta$ run through all 31 orbitals as we have introduced in the main text.
The kinetic matrix $t_{\alpha\beta}(\vk)$ only contains the elements corresponding to the \emph{hoppings} and \emph{hybridizations}. 
All terms that correspond to the on-site potentials are denoted as $\tilde{\varepsilon}_\alpha$, and are not included in $t_{\alpha\beta}(\vk)$. 
The interaction Hamiltonian will be chosen as the form shown in Eq.~(1) in the main text, in which the repulsive $U$ and the Hund's coupling $J_H$ only apply on the Cr-$d$ orbitals.

Each local fermionic operator $c^\dagger_{j\alpha\sigma}$ is mapped to a product of two parton operators:
\begin{equation}
    c^\dagger_{j\alpha\sigma} \rightarrow f^\dagger_{j\alpha\sigma}o^\dagger_{j\alpha\sigma}\,,
\end{equation}
Here, $f^\dagger_{j\alpha\sigma}$ is fermionic, and $o^\dagger_{j\alpha\sigma}$ is a spin-$\frac12$ operator, which acts on a local 2-dimensional Hilbert space. In the framework of $U(1)$ slave-spin theory, the spin operator $o^\dagger$ has the following form:
\begin{equation}
    o^\dagger_{j\alpha\sigma} = P^+_{j\alpha\sigma} S^+_{j\alpha\sigma}P^-_{j\alpha\sigma}\,,~~~P^\pm_{j\alpha\sigma} = \frac{1}{\sqrt{\frac{1}{2} \pm S^z_{j\alpha\sigma}}}\,,
\end{equation}

Because of this parton construction, the local Hilbert space dimension is increased from 2 to 4. A local constraint of the physical Hilbert space is required. We introduce Lagrange multipliers terms for each orbital into the parton Hamiltonian:
\begin{equation}
    H_\lambda = \sum_{j\alpha\sigma} \lambda_\alpha\left(S^z_{j\alpha\sigma} + \frac12 - f^\dagger_{j\alpha\sigma}f_{j\alpha\sigma}\right)\,,
\end{equation}
and the local constraint $\langle S^z_{j\alpha\sigma} \rangle + 1/2 = \langle f^\dagger_{j\alpha\sigma}f_{j\alpha\sigma} \rangle$ can be satisfied on the mean field level by controlling the values of $\lambda_\alpha$. 
The kinetic Hamiltonian can be written as the following form using these operators:
\begin{equation}\label{eqn:kinetic-parton}
    H_0 = \sum_{ij\alpha\beta\sigma}t_{\alpha\beta}^{ij}f^\dagger_{i\alpha\sigma}f_{j\beta\sigma} o^\dagger_{i\alpha\sigma}o_{j\beta\sigma} + \sum_{j\alpha\sigma}(\varepsilon_\alpha - \lambda_\alpha)f^\dagger_{j\alpha\sigma}f_{j\alpha\sigma}\,,
\end{equation}
in which $t_{\alpha\beta}^{ij}$ is the Fourier transformation of $t_{\alpha\beta}(\vk)$ represented in real space indices. The hopping terms can then be decomposed using a ``mean-field'' assumption and a ``single-site approximation'' for the slave-spin operators:
\begin{equation}    
    f^{\dagger}_{i\alpha\sigma}f_{j\beta\sigma}o^\dagger_{i\alpha\sigma}o_{j\beta\sigma} \rightarrow \langle o^\dagger_{i\alpha\sigma}\rangle \langle o_{j\beta\sigma} \rangle f^\dagger_{i\alpha\sigma}f_{j\beta\sigma} + \langle f^\dagger_{i\alpha\sigma}f_{j\beta\sigma}\rangle\left(\langle o^\dagger_{i\alpha\sigma} \rangle o_{j\beta\sigma} + o^\dagger_{i\alpha\sigma}\langle o_{j\beta\sigma} \rangle\right)\,.\label{eqn:kinetic-mean-field}
\end{equation}
We will then assume the ``condensate'' of the slave-spin operators correspond to a translation invariant non-magnetic state, such that $\langle o^\dagger_{i\alpha\sigma} \rangle = \langle o^\dagger_{i\alpha \bar{\sigma}}\rangle = \langle o^\dagger_{j\alpha\sigma} \rangle = \sqrt{Z_{\alpha}}$. A Taylor expansion can also be applied to the $o^\dagger$ operators, which can lead to a simpler form:
\begin{equation}
    o^\dagger_{j\alpha\sigma} \approx \frac{S^+_{j\alpha\sigma}}{\sqrt{n_\alpha(1 - n_\alpha)}} + \sqrt{Z_{\alpha}}\eta_\alpha\left[2S^z_{j\alpha\sigma} - (2n_\alpha - 1) \right],\label{eqn:mf-decoupling}
\end{equation}
in which $\eta_\alpha = \frac12 \frac{n_\alpha - 1/2}{n_\alpha(1 - n_\alpha)}$ and $n_\alpha = \langle f^\dagger_{j\alpha\sigma} f_{j\alpha\sigma} \rangle$ is the fermion density. Here the site and spin indices $j,\sigma$ are omitted in the notation $n_\alpha$, because we assume that the solution is translation invariant and non-magnetic.

Applying Eq.~(\ref{eqn:kinetic-mean-field}) to Eq.~(\ref{eqn:kinetic-parton}), the kinetic Hamiltonian can be separated into two parts, one of which only contains the slave-spin operators. Moreover, slave-spin operators on different lattice sites are decoupled from each other because of the single-site approximation in Eq.~(\ref{eqn:mf-decoupling}). 
Combining these terms with the local interaction terms, the Hamiltonian of the slave-spin operators can be written as follows:
\begin{equation}
    H^S = U\sum_{\alpha}S^z_{\alpha, \uparrow} S^z_{\alpha\downarrow}+ U' \sum_{\alpha\neq\beta}S^z_{\alpha\uparrow}S^z_{\beta\downarrow} + (U' - J_H)\sum_{\alpha < \beta, \sigma }S^z_{\alpha\sigma}S^z_{\beta\sigma} + \sum_{\alpha\sigma}\lambda_\alpha S^z_{\alpha\sigma} + \sum_{\alpha \sigma}\left[h_\alpha \frac{S^+_{\alpha\sigma}}{\sqrt{n_\alpha(1 - n_\alpha)}} + {\rm H.c.}\right]\,,
\end{equation}
in which the ``bath'' fields $h_\alpha$ are given by:
\begin{equation}
    h_\alpha = \frac{1}{N}\sum_{\mathbf{k}\beta}t_{\alpha\beta}(\vk)\sqrt{Z_\beta} \langle f^\dagger_{\vk\alpha\sigma} f_{\vk\beta\sigma} \rangle\,.
\end{equation}
Note that $H^S$ contains parameters that are determined by the expectation values of slave-fermion operators, such as $n_\alpha = \langle f^\dagger_{j\alpha\sigma} f_{j\alpha\sigma} \rangle$.
The other part of the mean-field decomposition of $H_0$, which only contains the quadratic terms of slave-fermion operators, will have parameters that are determined by the expectation of the slave-spin operators, such as $\sqrt{Z_\alpha} = \langle o^\dagger_{j\alpha\sigma} \rangle$. The slave-fermion Hamiltonian reads:
\begin{align}
    H^f &= \sum_{\vk\alpha\beta\sigma}h^f_{\alpha\beta}(\vk) f^\dagger_{\vk\alpha\sigma}f_{\vk \beta\sigma}\,,\\
    h^f_{\alpha\beta}(\vk) &= \sqrt{Z_\alpha Z_\beta}t_{\alpha\beta}(\vk) + \delta_{\alpha\beta}(\tilde{\varepsilon}_\alpha - \lambda_\alpha + \lambda^0_\alpha - E_F)\,.
\end{align}
in which $E_F$ is the Fermi level determined via the filling factor. The parameters $\lambda^0_\alpha$ are defined as:
\begin{equation}
    \lambda^0_\alpha = -\sqrt{Z_\alpha}|h_\alpha|\frac{2 n_\alpha - 1}{n_\alpha(1 - n_\alpha)}\,,
\end{equation}
which originate from the second term on the right-hand-side of Eq.~(\ref{eqn:mf-decoupling}). 
For a given total filling factor, these mean-field parameters $\sqrt{Z_\alpha}$, $n_\alpha$, $\lambda_\alpha$, $\lambda^0_\alpha$ and $E_F$ can all be solved self-consistently.

Eigenvalues of the slave-fermion Hamiltonian $h^f(\vk)$ provide the dispersion of the coherent excitations, in which $Z_\alpha = \langle o^\dagger_{j\alpha\sigma}\rangle \langle o_{j\alpha\sigma}\rangle$ stand for the quasiparticle weight, and $\Delta \tilde{\varepsilon}_\alpha = \lambda^0_\alpha - \lambda_\alpha$ stand for the on-site energy renormalization of orbital $\alpha$.
Note that only the Cr-$d$ orbitals have non-trivial $Z_\alpha$, $\lambda_\alpha$ and $\lambda_\alpha^0$ values. For the Sb-$p$ orbitals, we always have $Z_\alpha = 1$ and $\lambda_\alpha = \lambda^0_\alpha$. 
Consequently, the coherent part of the Green's function will have the following form:
\begin{equation}
    G^{\rm coh}_{\alpha\beta}(\vk, z) = \sqrt{Z_\alpha Z_\beta}\left[\omega - h^f(\vk)\right]^{-1}_{\alpha\beta}\,,
\end{equation}
and the corresponding momentum-resolved spectral function of orbital $\alpha$ can be written as:
\begin{equation}\label{eqn:def-Acoh}
    A^{\rm coh}_\alpha(\vk, \omega) = -\frac{1}{\pi}{\rm Im}\,G^{\rm coh}_{\alpha\alpha}(\vk, \omega + i0^+)\,.
\end{equation}
The dispersive spectral functions in Fig.~4(a) in the main text are computed using this equation. Similarly, the Fermi surface plots in Figs.~4(b-c) are also obtained from $A^{\rm coh}_\alpha(\vk, \omega=0)$.

The spectral function defined in Eq.~(\ref{eqn:def-Acoh}) only contains the contributions from the coherent parts, which correspond to the ground state of the slave-spin Hamiltonian $H^S$.
To compute the DOS of the Cr-$d$ orbitals, one needs to take the excited states of $H^S$ into consideration \cite{yu_mott_2011}:
\begin{align}
    A^{\rm loc}_\alpha(\omega) =& \frac{1}{N}\sum_{\vk}\sum_m \Bigg{(}\sum_{i, \epsilon_{\vk, i} > 0} \delta(\omega + E_g - E_m - \epsilon_{\vk,i}) |u_{\alpha, i}(\vk)|^2 \frac{|\langle m |S^+_\alpha|g\rangle|^2}{n_\alpha(1 - n_\alpha)}\mathcal{F}^+_{\alpha, m} \nonumber\\
    & + \sum_{i,\epsilon_{\vk, i} < 0}\delta(\omega - E_g + E_m - \epsilon_{\vk, i}) |u_{\alpha, i}(\vk)|^2 \frac{|\langle m |S^-_{\alpha}|g \rangle|^2}{n_\alpha(1 - n_\alpha)}\mathcal{F}^-_{\alpha, m}\Bigg{)}\,,\label{eqn:d-ldos}
\end{align}
in which $E_m$ and $|m\rangle$ stand for the eigenvalues and eigenstates of $H^S$. Specifically, $m = g$ represents the ground state. The dispersion relation $\epsilon_{\vk,i}$ and the ``Bloch state'' $u_{\alpha,i}(\vk)$ are defined as the eigenvalues and eigenvectors of $h^f(\vk)$.
The factors $\mathcal{F}^\pm_{\alpha, m}$ are given by the following expressions:
\begin{align}
    \mathcal{F}^+_{\alpha, m} &= \left\{ 
    \begin{array}{ll}
        1, & m = g,\\
        \frac{1 - Z_\alpha}{n^{-1}_\alpha - Z_\alpha}\,, & m \neq g, 
    \end{array}
    \right.\\
    \mathcal{F}^-_{\alpha, m} &= \left\{ 
    \begin{array}{ll}
        1, & m = g,\\
        \frac{1 - Z_\alpha}{(1-n_\alpha)^{-1} - Z_\alpha}\,, & m \neq g.
    \end{array}
    \right.
\end{align}
These factors $\mathcal{F}^\pm_{\alpha, m}$ guarantee the satisfaction of the spectral function sum rules.
For other orbitals, the DOS has the same expression as in non-interacting fermion systems, with the dispersion $\epsilon_{\vk, i}$ and Bloch states $u_{\alpha, i}(\vk)$ replaced by the eigenvalues and eigenvectors of the slave-spin Hamiltonian $h^f(\vk)$:
\begin{equation}
    A^{\rm loc}_{\alpha}(\omega) = \frac{1}{N} \sum_\vk \sum_{i}\delta(\omega - \epsilon_{vk,i}) |u_{\alpha, i}(\vk)|^2\,.\label{eqn:c-ldos}
\end{equation}
The DOS plots in Fig.~4(a) in the main text are computed using Eqs.~(\ref{eqn:d-ldos}) and (\ref{eqn:c-ldos}).

\section{Additional numerical results}\label{appsec:add-result}

\subsection{Band structure and \texorpdfstring{$d_{z^2}$}{dz2}, \texorpdfstring{$d_{yz}$}{dyz} and \texorpdfstring{$d_{xy}$}{dxy} orbitals}\label{appsec:fat-band}

\begin{table}
    \centering
    \begin{tabular}{c|c|c}
        \hline\hline
        orbital & $\mu_\alpha$ $(\rm eV)$ & $\sigma_\alpha$ $(\rm eV)$\\
        \hline
        $d_{z^2}$ &  $-0.935$ & $1.735$\\
        $d_{xz}$ & $-0.779$ & $1.474$ \\
        $d_{yz}$ & $-0.091$ & $1.903$\\
        $d_{x^2 - y^2}$ & $-0.190$ & $1.976$ \\
        $d_{xy}$ & $-0.816$ & $1.809$\\
        \hline\hline
    \end{tabular}
    \caption{The density of states distribution of the $d$ orbitals.}
    \label{tab:dorb-distribution}
\end{table}

\begin{figure}
    \centering
    \includegraphics[width=0.8\linewidth]{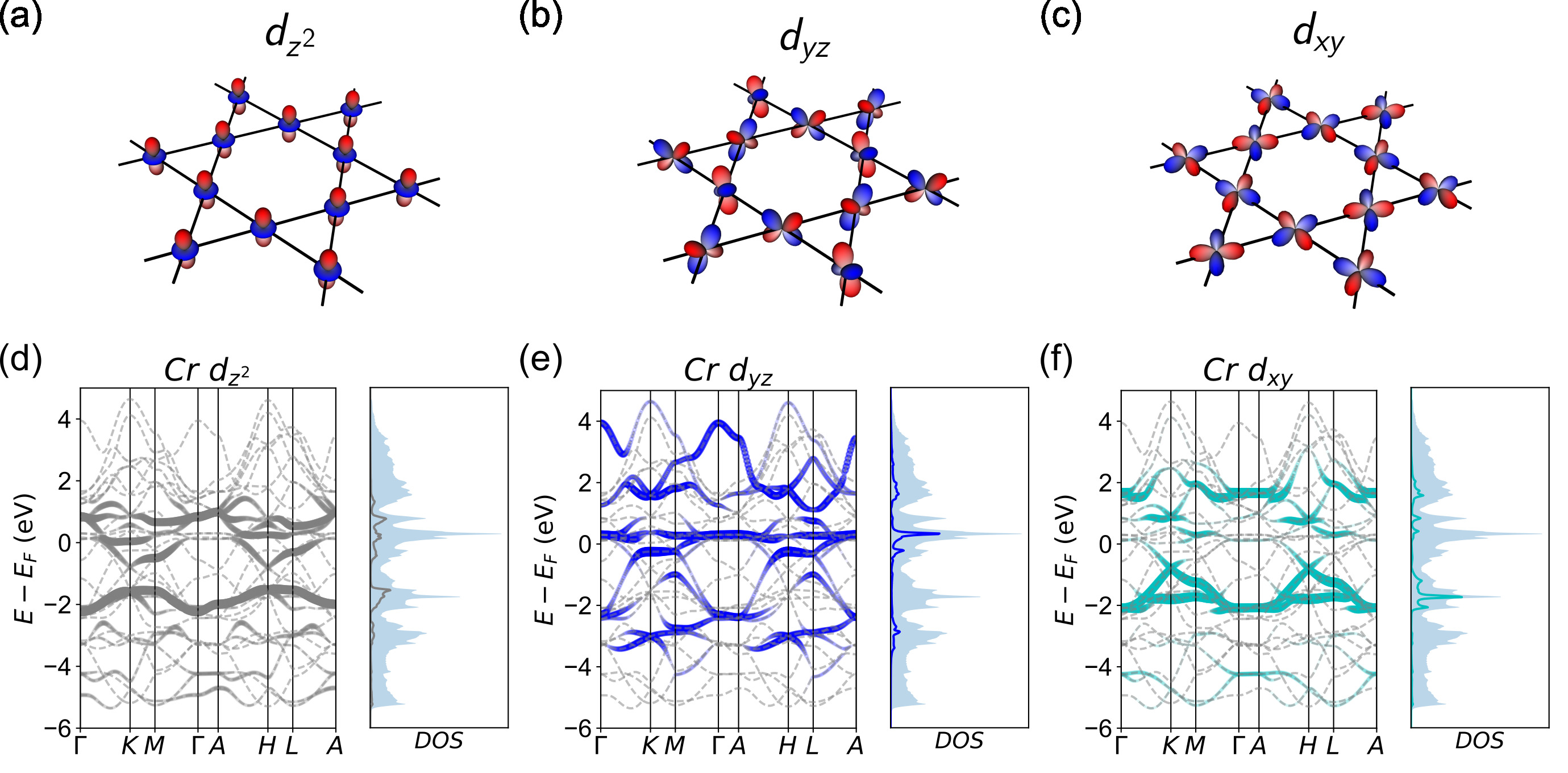}
    \caption{(a-c) The shape and orientation of the $d_{z^2}$, $d_{yz}$ and $d_{xy}$ orbitals from the chromium atoms. (d-f) The band structures and DOS projected onto these $d$ orbitals. Similar to Fig.~2, the light blue shadows stand for the total DOS.}
    \label{fig:orbs_app}
\end{figure}

In the main text, we provided the quasiparticle weights of the $d_{xz}$ and $d_{x^2 - y^2}$ orbitals in Fig.~2. Here in Fig.~\ref{fig:orbs_app}, we show the weights and the corresponding DOS of the other three types of $d$ orbitals.
Similar to the $d_{x^2 - y^2}$ orbital, we note that the $d_{yz}$ orbital also has a broader energy spread than other orbitals.

We note that the DOS of the different $d$ orbitals are distributed in different energy ``band widths'', which effectively capture the strength of hybridization and hopping of the corresponding orbital.
To quantify their effective ``band widths'', the following two quantities can be defined for each Cr-$d$ orbital:
\begin{align}
    \mu_\alpha &= \int_{-\infty}^\infty d\omega\,\omega A_\alpha(\omega)\,,\\
    \sigma_\alpha^2 &= \int_{-\infty}^\infty d\omega\,(\omega - \mu_\alpha)^2A_\alpha(\omega)\,,
\end{align}
in which $A_\alpha(\omega)$ stands for the density of states (spectral function) of the orbital $\alpha$. $\mu_\alpha$ and $\sigma_\alpha$ describe the ``average energy'' and ``band width'' of the corresponding orbital. In Table \ref{tab:dorb-distribution}, the values of these two quantities for the Cr-$d$ orbitals are presented. We found that the $d_{xz}$ has the smallest value of $\sigma_\alpha$, while the $d_{yz}$ and $d_{x^2 - y^2}$ have larger $\sigma_\alpha$ values. 
One can also notice that all of these five $d$ orbitals have non-negligible contributions to the electronic states near the Fermi level, suggesting that a truncated effective model with fewer Cr-$d$  orbitals is likely inadequate.

\begin{figure}
    \centering
    \includegraphics[width=\linewidth]{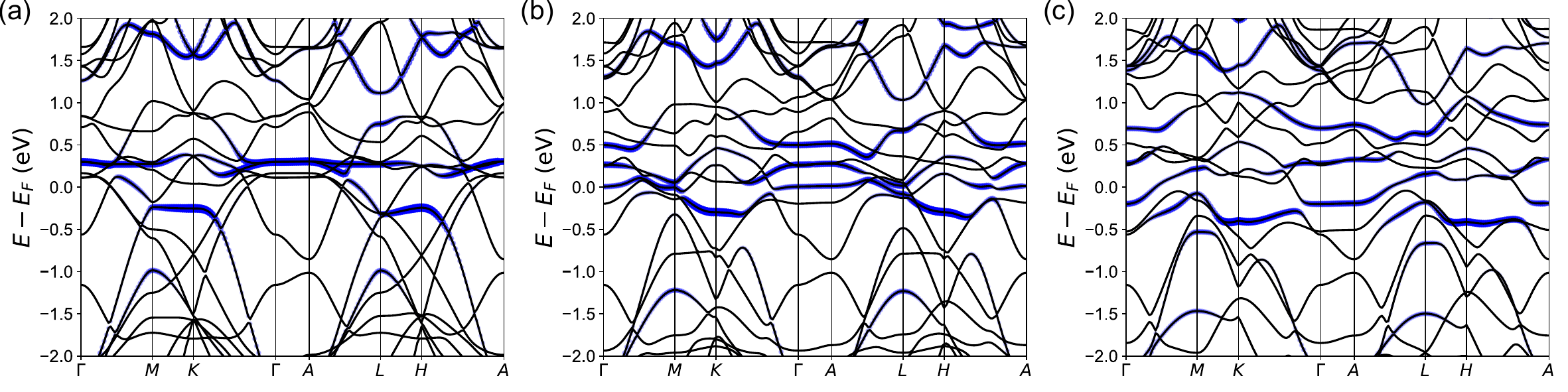}
    \caption{(a) The band structure of the $31$-orbital tight-binding model for CsCr$_3$Sb$_5$.
    (b) The band structure of the ``distorted'' tight-binding model with on-site energy shift $\delta = 0.5 \rm \, eV$.
    (c) The band structure of the ``distorted'' tight-binding model with on-site energy shift $\delta = 1\rm\, eV$.
    Here the Cr-$d_{yz}$ orbital content is highlighted in blue.}
    \label{fig:distortion}
\end{figure}

We also note that the value of $\sigma_\alpha$ for all five Cr-$d$ orbitals are on the order of $\rm eV$, significantly exceeding the width of the narrow bands near $E_F$.
Such observation indicates that the flatness of these bands are {\it not} because of the weak hoppings among Cr-$d$ and Sb-$p$ orbitals.
To further demonstrate the geometric nature of the narrow bands around $E_F$, we performed a band structure calculation in the presence of a ``distortion".
We do so by incorporating an on-site energy shift for two of the three Cr atoms per unit cell.
In particular, we add a positive energy shift $\delta$ to all of the five $d$ orbitals of one of the Cr atom, and a negative energy shift $-\delta$ to all of the five $d$ orbitals of another Cr atom.
The results are shown in Fig.~\ref{fig:distortion}. 
With this distortion taken into account, the effect of geometric frustration is expected to be effectively reduced, as evidenced by the noticeably increased bandwidths of the narrow bands.

\subsection{Electronic structure with different interaction strength}\label{appsec:add-electronic-structure}

\begin{figure}
    \centering
    \includegraphics[width=\linewidth]{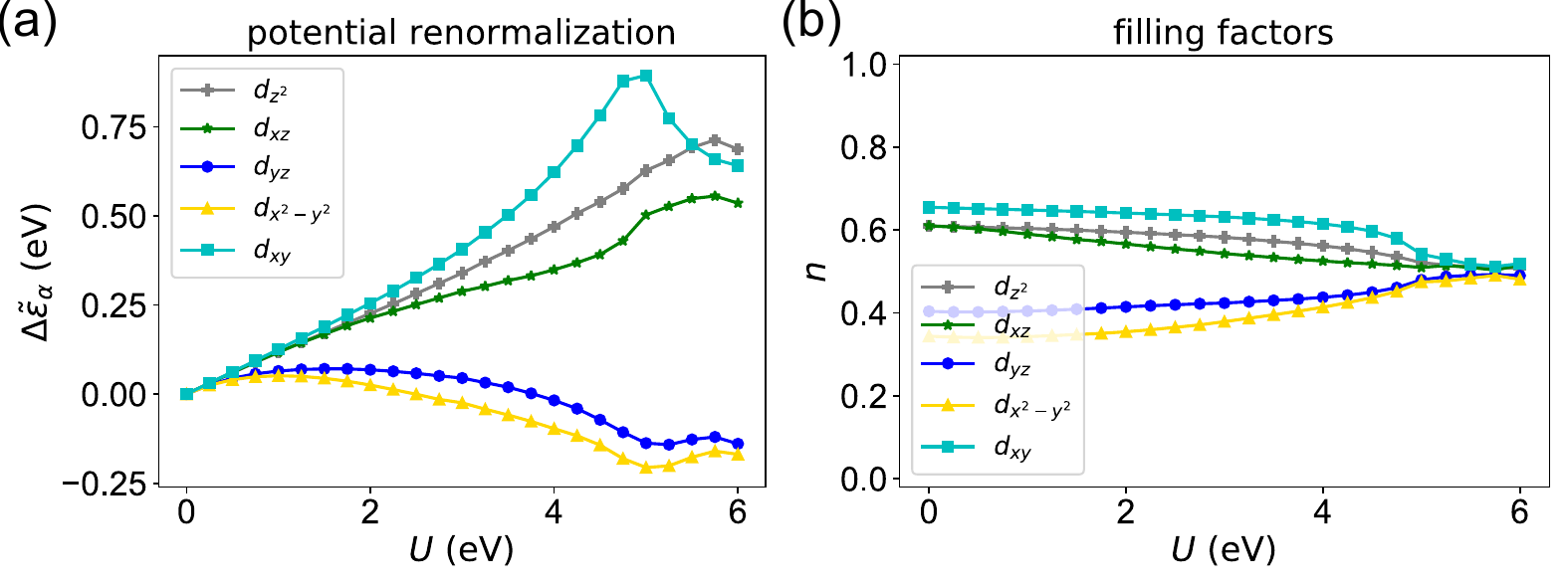}
    \caption{
    (a) The on-site potential renormalization $\Delta \varepsilon_\alpha = \lambda^0_\alpha - \lambda_\alpha$ of each Cr-$d$ orbital as functions of $U$.
    (b) Average particle numbers of each Cr-$d$ orbital as functions of interaction strength $U$.
    }
    \label{fig:n-lam-renormalization}
\end{figure}

As we have discussed in the main text, the quasiparticle weights of the Cr-$d$ orbitals are suppressed by the interaction effects. In the mean time, the on-site potential energies and the relative filling factors of these Cr-$d$ orbitals are also affected. In Fig.~\ref{fig:n-lam-renormalization}(a), we show the $U$-dependency of the on-site energy renormalization $\Delta \tilde{\varepsilon}_\alpha = \lambda^0_\alpha - \lambda_\alpha$. Qualitatively speaking, the on-site potentials of the $d_{yz}$ and $d_{x^2 - y^2}$ orbitals are pushed downward, while the other three Cr-$d$ orbitals are pushed upward. This is compatible with the filling factors change as shown in Fig.~\ref{fig:n-lam-renormalization}(b), in which all the five Cr-$d$ orbitals filling factors are renormalized to half-filling in the strong interaction limit. When the local interaction $U$ is turned off, the $d_{yz}$ and $d_{x^2 - y^2}$ orbitals are below half filling, and a negative on-site energy renormalization is able to increase the filling factors of the corresponding orbitals.

\begin{figure}
    \centering
    \includegraphics[width=\linewidth]{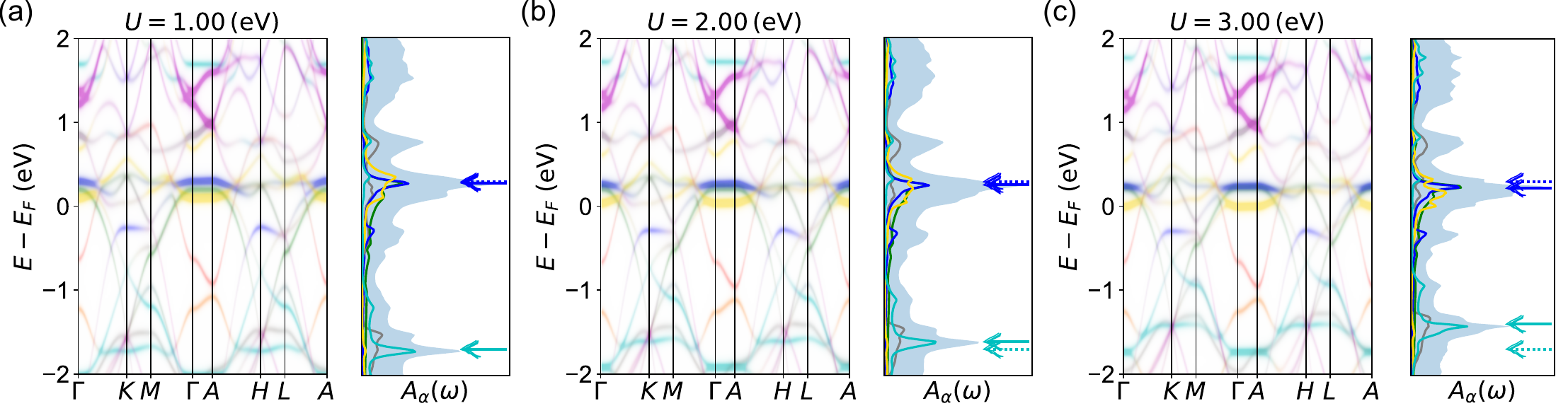}
    \caption{The coherent excitation dispersions and orbital resolved DOS with interaction strengths (a) $U = 1\,\rm eV$, (b) $U = 2\,\rm eV$ and (c) $U = 3 \,\rm eV$.
    Here the dashed arrows in the DOS plots indicate the peaks corresponding to the DFT band structure, and the solid arrows label the peaks corresponding to the correlated band structure with different $U$ values.
    }
    \label{fig:ss-bands-app3}
\end{figure}

In Fig.~\ref{fig:ss-bands-app3}, we provide the correlated electronic structure, including the coherent excitation dispersions and the orbital resolved DOS plots with weak on-site interaction strength $U = 1{\rm\, eV}, 2{\rm\, eV}$ and $3{\rm\, eV}$.
In the DOS plots, the dashed arrows label the position of the spectral peaks of the {\it ab initio} band structure without considering the strong electronic correlations, and the solid arrows stand for the position of the spectral peaks with correlation effects.
This analysis clearly shows that the correlated flat bands pinned near $E_F$, as showin in Fig.\,4(a) of the main text, are adiabatically connected to the destructive-interference-induced DFT flat bands above $E_F$ in the noninteracting case; in other words, the correlated flat bands pinned near $E_F$ result from the combined effects of destructive interference and electron correlations.

\begin{figure}
    \centering
    \includegraphics[width=\linewidth]{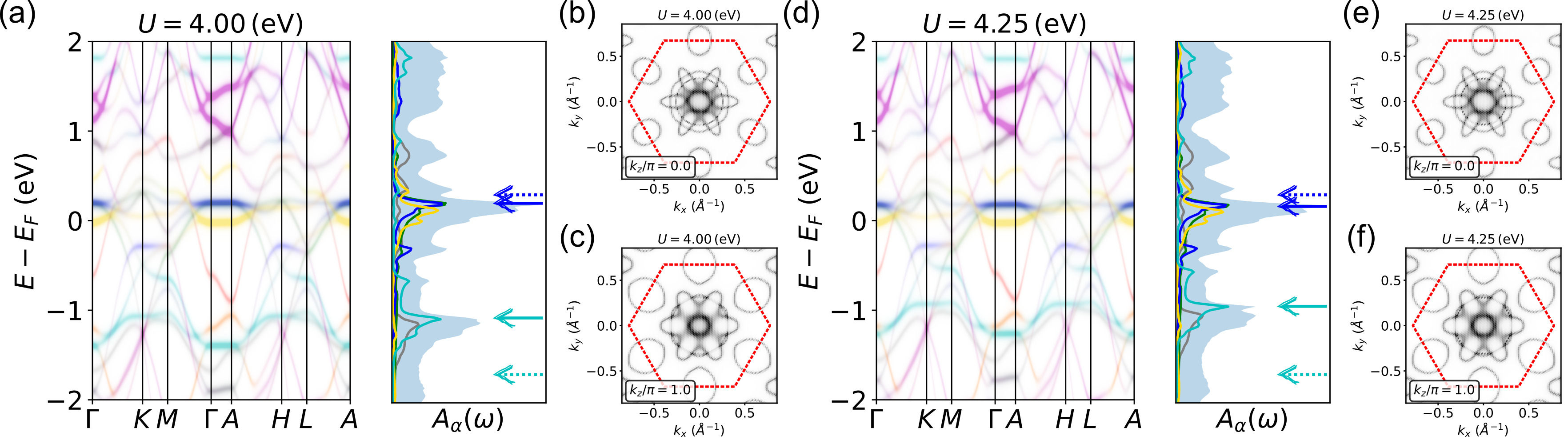}
    \caption{(a-c) The coherent excitation dispersions, orbital-resolved DOS and Fermi surfaces at $k_z = 0$ and $k_z = \pi$ with interaction strength $U = 4.0\rm\,eV$. 
    (d-f) The coherent dispersions, DOS and Fermi surfaces solved with interaction strengths $U = 4.25\,\rm eV$.
    }
    \label{fig:ss-bands-app1}
\end{figure}

\begin{figure}
    \centering
    \includegraphics[width=\linewidth]{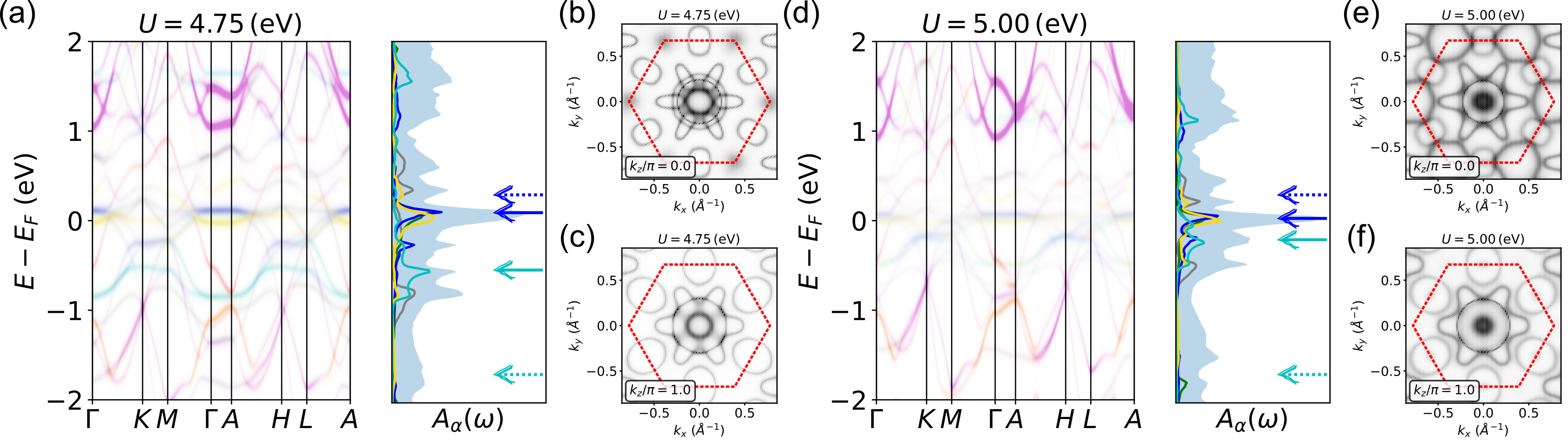}
    \caption{(a-c) The coherent excitation dispersions, the orbital-resolved DOS and the Fermi surfaces at $k_z = 0$ and $k_z = \pi$ with the interaction strength $U = 4.75\rm\,eV$. 
    (d-f) The coherent dispersions, the DOS and the Fermi surfaces solved with 
    the interaction strengths $U = 5\,\rm eV$.
    }
    \label{fig:ss-bands-app2}
\end{figure}

We also provide the electronic structure, including the coherent spectral functions, the DOS, and the Fermi surface structures with larger interaction strength in Figs.~\ref{fig:ss-bands-app1} and \ref{fig:ss-bands-app2}. The spectral functions shown in Fig.~\ref{fig:ss-bands-app1}(a) are computed with interaction strength $U = 4\,\rm eV$. 
With this interaction strength, the quasiparticle weights are renormalized to around $Z \gtrsim 0.7$, and the coherent excitation dispersion is already modified. 
The solid arrows, which indicate the peaks of the DOS, are dragged closer to the Fermi level. 
Similar to the case discussed in the main text, the size and shape of the concentric Fermi pockets around $\Gamma$ point are also changed from the ones at $U = 0$.
In the case with $U = 4.25\,\rm eV$ shown in Figs.~\ref{fig:ss-bands-app1}(d-f), similar but slightly stronger correlation effects can be observed.

When the interaction is further increased to $U = 4.75\rm\, eV$, the renormalization effects shown in Fig.~\ref{fig:ss-bands-app2}(a) are even stronger. The quasiparticle weights are reduced to $Z \sim 0.2$, and stronger modification to the band structure can be seen. Besides, as shown in Fig.~\ref{fig:ss-bands-app2}(b), the increased intensity near $K$ and $K'$ points, indicating the development of new Fermi pockets, which is another signal of modified electronic structure.

In Fig.~\ref{fig:ss-bands-app2}(d), we show the coherent spectral functions with interaction strength $U = 5\rm\,eV$.
Clearly, the DOS peak labeled by the solid blue arrow has already been pinned extremely close to $E_F$, showing the possibility of an emergent flat band.
The other DOS peak, which is mostly formed by $d_{xy}$ orbital and was around $2\,\rm eV$ below $E_F$ in the {\it ab initio} band structure, is also pushed within $\lesssim 0.5\rm\,eV$ from $E_F$.
In addition, the Fermi surfaces, which are shown in Fig.~\ref{fig:ss-bands-app2}(e-f), are significantly changed with interaction strength $U = 5\rm\,eV$. Except for the size and shape changes of the pockets around $\Gamma$ and $M$, there are indeed new electron pockets around $K$ and $K'$ points developed. As such, this material can potentially be tuned through a Lifshitz transition.

\subsection{Band structure with spin-orbit coupling}

\begin{figure}
    \centering
    \includegraphics[width=0.5\linewidth]{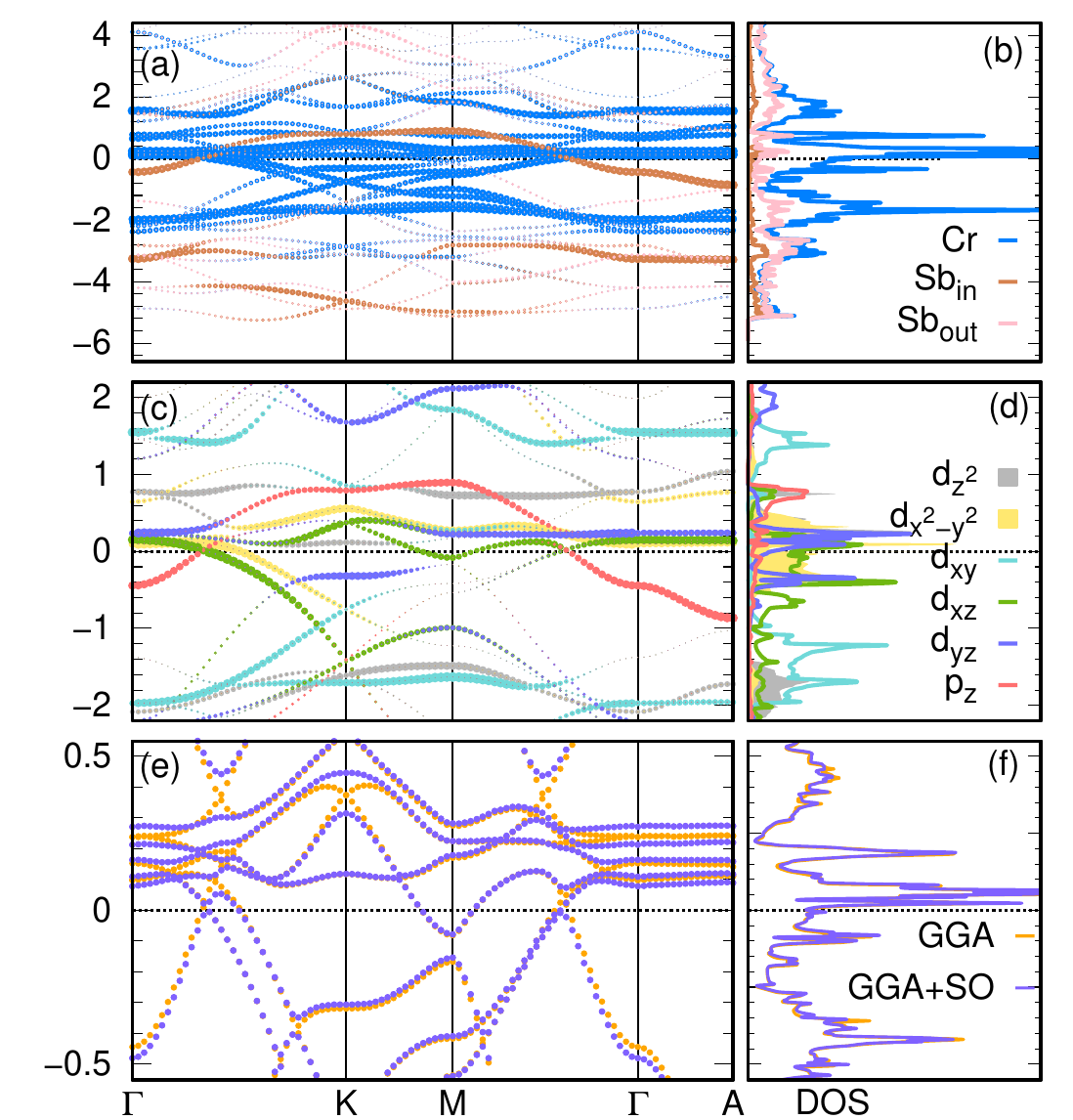}
    \caption{(a-b) The overall band structure of $\rm CsCr_3 Sb_5$.
    (c-d) The Cr-$d$ and Sb-$p$ orbitals projected onto the band structure in the energy interval $-2{\rm\,eV} \leq E - E_F \leq 2 \rm\,eV$.
    (e-f) The band structure with and without spin-orbit coupling in the energy interval $-0.5{\rm\,eV} \leq E - E_F \leq 0.5{\rm \, eV}$.}
    \label{fig:dos-soc}
\end{figure}

In Fig.~\ref{fig:dos-soc}, we provide the plots of the {\it ab initio} band structure with and without the spin-orbit coupling. As seen in Fig.~\ref{fig:dos-soc}(e), SOC could lead to some gap opening up to $\sim 0.1{\rm\, eV}$ at $\Gamma$ and $K$ points.

\end{document}